\documentclass[final,3p,times,twocolumn]{elsarticle}

\usepackage{amssymb}

\usepackage{subfig}
\usepackage{amsmath}
\usepackage{multirow}
\usepackage{amsfonts}       
\usepackage{diagbox}
\usepackage{booktabs}
\usepackage{algorithm}
\usepackage[noend]{algpseudocode}

\makeatletter
\def\BState{\State\hskip-\ALG@thistlm}
\makeatother

\journal{Neurocomputing}

\begin{document}

\begin{frontmatter}



\title{Learning based Facial Image Compression with Semantic Fidelity Metric}


\author{Zhibo Chen}
\ead{chenzhibo@ustc.edu.cn}
\author{Tianyu He}
\address{CAS Key Laboratory of Technology in \\ Geo-spatial Information Processing and Application System \\ University of Science and Technology of China, Hefei, China}

\begin{abstract}
Surveillance and security scenarios usually require high efficient facial image compression scheme for face recognition and identification.  While either traditional general image codecs or special facial image compression schemes only heuristically refine codec separately according to face verification accuracy metric. We propose a Learning based Facial Image Compression (LFIC) framework with a novel Regionally Adaptive Pooling (RAP) module whose parameters can be automatically optimized according to gradient feedback from an integrated hybrid semantic fidelity metric, including a successfully exploration to apply Generative Adversarial Network (GAN) as metric directly in image compression scheme. The experimental results verify the framework's efficiency by demonstrating performance improvement of 71.41\%, 48.28\% and 52.67\% bitrate saving separately over JPEG2000, WebP and neural network-based codecs under the same face verification accuracy distortion metric. We also evaluate LFIC's superior performance gain compared with latest specific facial image codecs. Visual experiments also show some interesting insight on how LFIC can automatically capture the information in critical areas based on semantic distortion metrics for optimized compression, which is quite different from the heuristic way of optimization in traditional image compression algorithms.
\end{abstract}

\begin{keyword}
end-to-end \sep semantic metric \sep facial image compression \sep adversarial training



\end{keyword}

\end{frontmatter}


\section{Introduction}
\label{introduction}

\begin{figure}[t]
	\centerline{\includegraphics[width=0.99\linewidth]{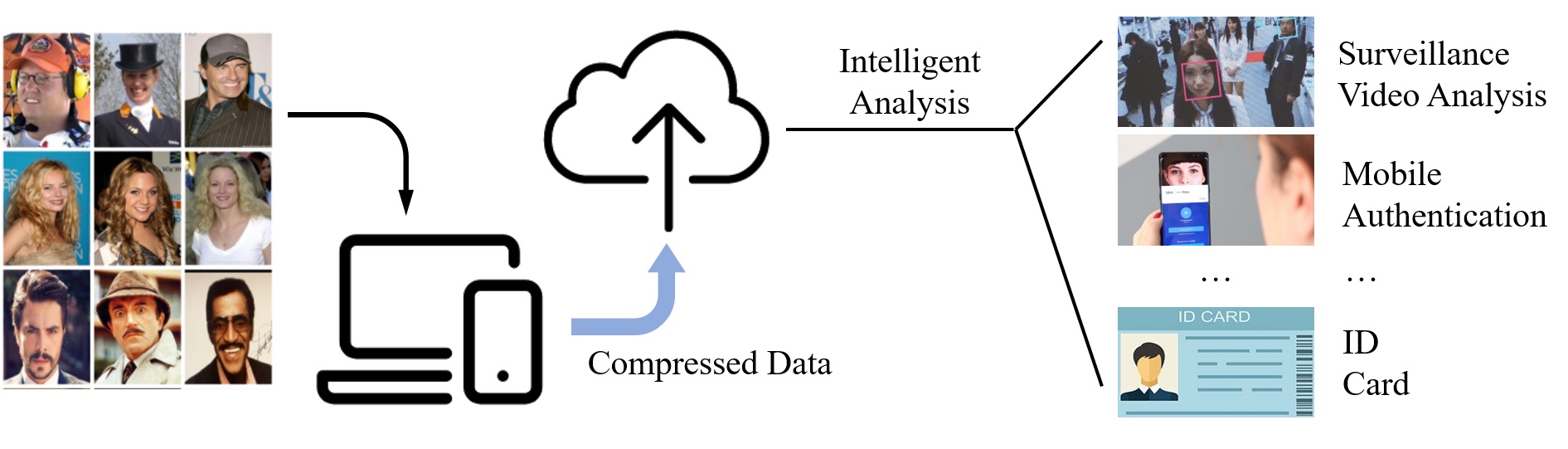}}
	\caption{A highly efficient facial image compression is broadly required (located in blue arrow) in a wide range of intelligent applications.}
	\centering
	\label{fig:fig1}
\end{figure}

Face verification/recognition has been developing rapidly in recent years, which facilitates a wide range of intelligent applications such as surveillance video analysis, mobile authentication, etc. Since these frequently-used applications generate a huge amount of data that requires to be transmitted or stored, a highly efficient facial image compression is broadly required as illustrated in Fig. \ref{fig:fig1}.

Basically, facial image compression can be regarded as a special application of general image compression technology. While evolution of general image/video compression techniques has been focused on continuously improving Rate Distortion performance, viz. reducing the compressed bit rate under the same distortion between the reconstructed pixels and original pixels or reducing the distortion under the same bit rate. The apparent question is how to define the distortion metric, especially for specific application scenario such as face recognition in surveillance. Usually we can classify the distortion into three levels of distortion metrics: \textbf{Pixel Fidelity}, \textbf{Perceptual Fidelity}, and \textbf{Semantic Fidelity}, according to different levels of human cognition on image/video signals.

The most common metric is \textbf{Pixel Fidelity}, which measures the pixel level difference between the original image and compressed image, \textit{e.g.}, MSE (Mean Square Error) has been widely adopted in many existed image and video coding techniques and standards (\textit{e.g.}, MPEG-2, H.264, HEVC, etc.). It can be easily integrated into image/video hybrid compression framework as an in-loop metric for rate-distortion optimized compression. However, it's obvious that pixel fidelity metric cannot fully reflect human perceptual viewing experience \cite{wan2009mean}. Therefore, many researchers have developed \textbf{Perceptual Fidelity} metrics to investigate objective metrics measuring human subjective viewing experience \cite{chen2016hybrid}. With the development of aforementioned intelligent applications, image/video signals will be captured and processed for semantic analysis. Consequently, there will be increasingly more requirements on research for \textbf{Semantic Fidelity} metric to study the semantic difference (\textit{e.g.}, difference of verification accuracy) between the original image and compressed image. There are few research work on this area \cite{chopra2005learning,zhang2015som} and usually are task-specific.

The aforementioned various distortion metrics provide a criteria to measure the quality of reconstructed content. However, the ultimate target of image quality assessment is not only to measure the quality of images with different level of distortion, but also to apply these metrics to optimize image compression schemes. While it's a contradictory that most complicated quality metrics with high performance are not able to be integrated easily into an image compression loop. Some research works tried to do this by adjusting image compression parameters (\textit{e.g.}, Quantization parameters) heuristically according to embedded quality metrics \cite{alakuijala2017guetzli,liu2017recognizable}, but they are still not fully automatic-optimized end-to-end image encoder with integration of complicated distortion metrics.

\begin{figure}[t]
	\centering
	\subfloat[]{\includegraphics[width=0.22\textwidth]{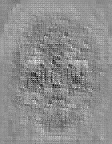}%
		\label{fig:fig2a}}
	\hfil
	\subfloat[]{\includegraphics[width=0.22\textwidth]{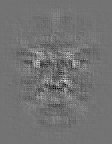}%
		\label{fig:fig2b}}
	\caption{The visualization of gradient feedback from (a) MSE; (b) the integrated face verification metric, which shows that more focus is on the distinguishable regions (\textit{e.g.}, eye, nose) according to such semantic distortion metric.}
	\label{fig:fig2}
\end{figure}

In this paper, we are trying to solve this problem by developing a Learning based Facial Image Compression (LFIC) framework, to make it feasible to automatically optimize coding parameters according to gradient feedback from the integrated hybrid facial image distortion metric calculation module. Different from traditional hybrid coding framework with prediction, transform, quantization and entropy coding modules, we separate these different modules inside or outside the end-to-end loop according to their derivable property. We demonstrated the efficiency of this framework with the simplest prediction, quantization and entropy coding module. We propose a new module called Regionally Adaptive Pooling (RAP) inside the end-to-end loop to improve the ability to configure compression performance. RAP has advantages of being able to control bit allocation according to distortion metrics' feedback under a given bit budget. Face verification accuracy is adopted as one semantic distortion metric to be integrated into LFIC framework.

Although we adopt the simplest prediction, quantization and entropy coding module, the LFIC framework has shown great improvement over traditional codec like JPEG2000, WebP and neural network-based codecs, and also demonstrates much better performance compared with existing specific facial image compression schemes. The visualization as illustrated in Fig. \ref{fig:fig2} shows that more focus is on the distinguishable regions (\textit{e.g.}, eye, nose) according to face verification metric. Also, it demonstrates that our LFIC framework can automatically capture the information in critical areas based on semantic distortion metric.

In general, our contributions are four-folds: 1) a Learning based Facial Image Compression framework; 2) a novel pooling strategy called RAP; 3) a successful exploration to apply Generative Adversarial Network (GAN) as metric to compression directly; 4) a starting exploration for semantic based image compression.

\section{Related Work}
\label{related_work}

\subsection{Image Compression}

For image compression, standard image codecs such as JPEG, JPEG2000 and WebP have been widely used , which have made remarkable achievements in general applications over the past few decades. However, such compression schemes are becoming increasingly difficult to meet the needs for advanced applications of semantic analysis. There are some preliminary heuristic explorations such as Alakuijala \textit{et al.} adopted the distortion metric from the perspective of perceptual-level to analogically optimize the JPEG encoder \cite{alakuijala2017guetzli}. And Prakash \textit{et al.} enhanced JPEG encoder by highlighting semantically-salient regions \cite{prakash2017semantic}.

There are also some face-specific image compression schemes proposed in academic area, some attempts \cite{elad2007low,bryt2008compression,ram2014facial,ferdowsi2015sparse} have been made to design dictionary-based coding schemes on this specific image type. Moreover, face verification in compressed domain is another solution due to its lower computational complexity\cite{delac2008image}.

Recently image compression with neural network attracts increasing interest recently. Ball\'e \textit{et al.} optimized a model consisting of nonlinear transformations for a perceptual metric \cite{balle2016end1} and MSE \cite{balle2016end2}, also relaxed the discontinuous quantization with additive uniform noise with the goal of end-to-end training. Theis \textit{et al.} \cite{theis2017lossy} used a similar architecture but dealt with quantization and entropy rate estimation in a different way. Consistent with the architecture of \cite{theis2017lossy}, Agustsson \textit{et al.} \cite{agustsson2017soft} trained with a soft-to-hard entropy minimization scheme in the context of model and feature compression. Dumas \textit{et al.} \cite{dumas2017image} introduced a competition mechanism between image patches binding to sparse representation. Li \textit{et al.} \cite{li2017learning} achieved spatially bit-allocation for image compression by introducing an importance map. Jiang \textit{et al.} \cite{jiang2017end} realized a super-resolution-based image compression algorithm. As variable rate encoding is a fundamental requirement for compression, some efforts \cite{toderici2015variable,toderici2016full,johnston2017improved} have been devoted towards using autoencoders in a progressive manner, growing with the number of recurrent iterations. On the basis of these progressive autoencoders, Baig \textit{et al.} introduced an inpainting scheme that exploits spatial coherence to reduce redundancy in image \cite{baig2017learning}. Chen \textit{et al.} proposed an end-to-end framework for video compression \cite{chen2019learning}.
With the rapid development of GANs, it has been proved that it is possible to adversarially generate images from a compact representation \cite{santurkar2017generative,rippel2017real}.
In the last few months, the modeling of latent representation becomes an emerging direction \cite{balle2018variational,mentzer2018conditional}. Typically, they learned a probability model of the latent distribution to improve the efficiency of entropy coding.

However, most of aforementioned works either employed pixel fidelity and perceptual fidelity metrics, or optimized by heuristically adjusting codec's parameters. Instead, our framework is a neural network based scheme able to automatically optimize coding parameters with integrated hybrid distortion metrics, which demonstrates much higher performance improvement compared with these state of the art solutions. 

\subsection{Adaptive Pooling}

Traditional block based pooling strategy applied in neural network based scheme is not suitable for integrated semantic metrics, since most semantic metrics are not block-wise, e.g. face verification accuracy metric is to define the verification accuracy of the whole facial image rather than to define the accuracy of each block in the facial image. Therefore we need to propose a new pooling operation able to deal with this issue.

The idea of spatial pooling is to produce informative statistics in a specific spatial area. In consideration of relatively fixed pattern it has, several works aimed at enhancing its flexibility. Some approaches adaptively learned regions that distinguishable for classification \cite{jia2012beyond,he2014spatial}. Similar to Jia \textit{et al.}, some works tried to design better spatial regions for pooling to reduce the effect of background noise with the goal of image classification \cite{liu2016adaptive,wang2016csps} and object detection \cite{tsai2015adaptive}. As traditional pooling operation adopt a fixed block size for each image, we propose a variable block size pooling scheme named RAP, which is configurable optimized on the basis of integrated distortion metrics and provide ability of preserving higher quality to crucial local areas.

\begin{figure*}[t]
	\centerline{\includegraphics[width=0.98\textwidth]{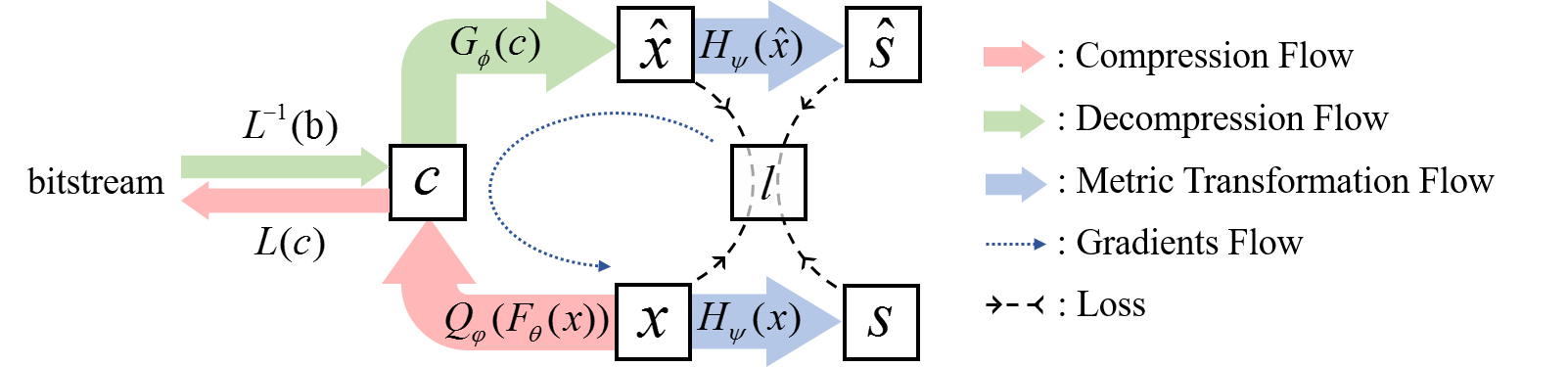}}
	\caption{The proposed Learning based Facial Image Compression (LFIC) Framework.}
	\centering
	\label{fig:fig3}
\end{figure*}

\section{Learning based Facial Image Compression Framework}
\label{compressionframework}

This section introduces the general framework of facial image compression with integrated general distortion metric, as illuminated in Fig. \ref{fig:fig3}.

\textbf{Compression Flow.} Consistent with conventional codec, our LFIC framework contains a compression flow and a decompression flow. In the compression flow, an image $x$ is fed into a differentiable encoder $F_\theta$ and a quantizer $Q_\varphi$, translated into a compact representation $c$: $c={Q_\varphi}(F_\theta(x))$, 
where subscripts refer to parameters (the same in the remaining of this section). The quantizer can attain a significant amount of data reduction, but still statistically redundant. Therefore, we further perform several generic or specific lossless compression schemes (\textit{i.e.}, transformation, prediction, entropy coding), formulated as $L(c)$, to achieve higher coding efficiency. After the lossless compression, $x$ is encoded into a bitstream $b$ that can be directly delivered to a storage device, or a dedicated link, etc.

\textbf{Decompression Flow.} In the decompression flow, due to the reversibility of lossless compression, $c$ can be recovered from the channel by $L^{-1}$. The reconstructed image $\hat{x}$ is ultimately obtained by a differentiable decoder $G_\phi$: $\hat{x}=G_\phi(c)$,
where $c=L^{-1}(b)$.

\textbf{Metric Transformation Flow.} As mentioned before, a general distortion metric calculation module is integrated into our LFIC framework. This motivates the use of a transformation $H_\psi$, that bridge the gap between pixel domain and metric domain. We expect that, the difference between $s$ and $\hat{s}$, which are generated from $x$ and $\hat{x}$ respectively, represents distortion measured in our desired metric domain (\textit{i.e.}, pixel fidelity domain, perceptual fidelity domain, and semantic fidelity domain). After that, the loss $l$ can be propagated back to each component of the compression-decompression flow (\textit{i.e.} $F_\theta$, $Q_\varphi$ and $G_\phi$) which needs to be differentiable.

\textbf{Gradients Flow.} Since $F_\theta$ and $G_\phi$ are both differentiable, therefore, the only inherently non-differentiable step here is quantization, which poses an undesirable obstacle for end-to-end optimization with gradient-based techniques. Some effective algorithms have been developed to tackle this challenging problem \cite{balle2016end1,toderici2015variable}. We follow the works in \cite{theis2017lossy}, which regards quantization as rounding, by replacing its derivative in backpropagation:
\begin{equation}\label{3}
\frac{d}{dF_\theta(x)}Q_\varphi(F_\theta(x)) := \frac{d}{dF_\theta(x)}[F_\theta(x)] := \frac{d}{dF_\theta(x)}R(F_\theta(x)),
\end{equation}
where $R$ is a smooth approximation, and square bracket depicts rounding a real number to the nearest integer value. We set $R(F_\theta(x))=F_\theta(x)$ here, which means perform backpropagation without modification through rounding. In general, the gradient of loss with respect to input image $x$ can be formulated as:
\begin{equation}\label{4}
\frac{\partial l}{\partial x} = \frac{\partial l}{\partial G_\phi} \frac{\partial G_\phi}{\partial Q_\varphi} \frac{\partial Q_\varphi}{\partial F_\theta} \frac{\partial F_\theta}{\partial x},
\end{equation}

In a word, during the entire pipeline of our framework, we separate distinct modules inside or outside the end-to-end loop according to their differentiable property. The modules like $F_\theta$, $Q_\varphi$, $G_\phi$ and $H_\psi$ are placed inside the loop, the parameters in these modules can be updated according to the gradient back-propagation from the loss measured by the distortion metric. The modules like $L$ and ${L^{-1}}$ are placed outside the loop, since it is non-differentiable and reversible.

\section{Semantic-oriented Facial Image Compression}
\label{facecompression}
As described in Sec. \ref{compressionframework}, compared with ordinary compression pipeline, the main advantage of semantic-oriented compression is that we can automatically optimize parameters in encoder $F_\theta$, quantizer $Q_\varphi$ and decoder $G_\phi$ according to the semantic distortion metric supported by $H_\psi$. It is significant that, we can preserve semantic features while reducing redundancy in a full-automatic way rather than heuristically tuning coding parameters according to distortion metrics, which is very important for future intelligent media applications with necessary of transmitting images compressed by preserving semantic fidelity.

\begin{figure*}[t]
	\centerline{\includegraphics[width=0.95\linewidth]{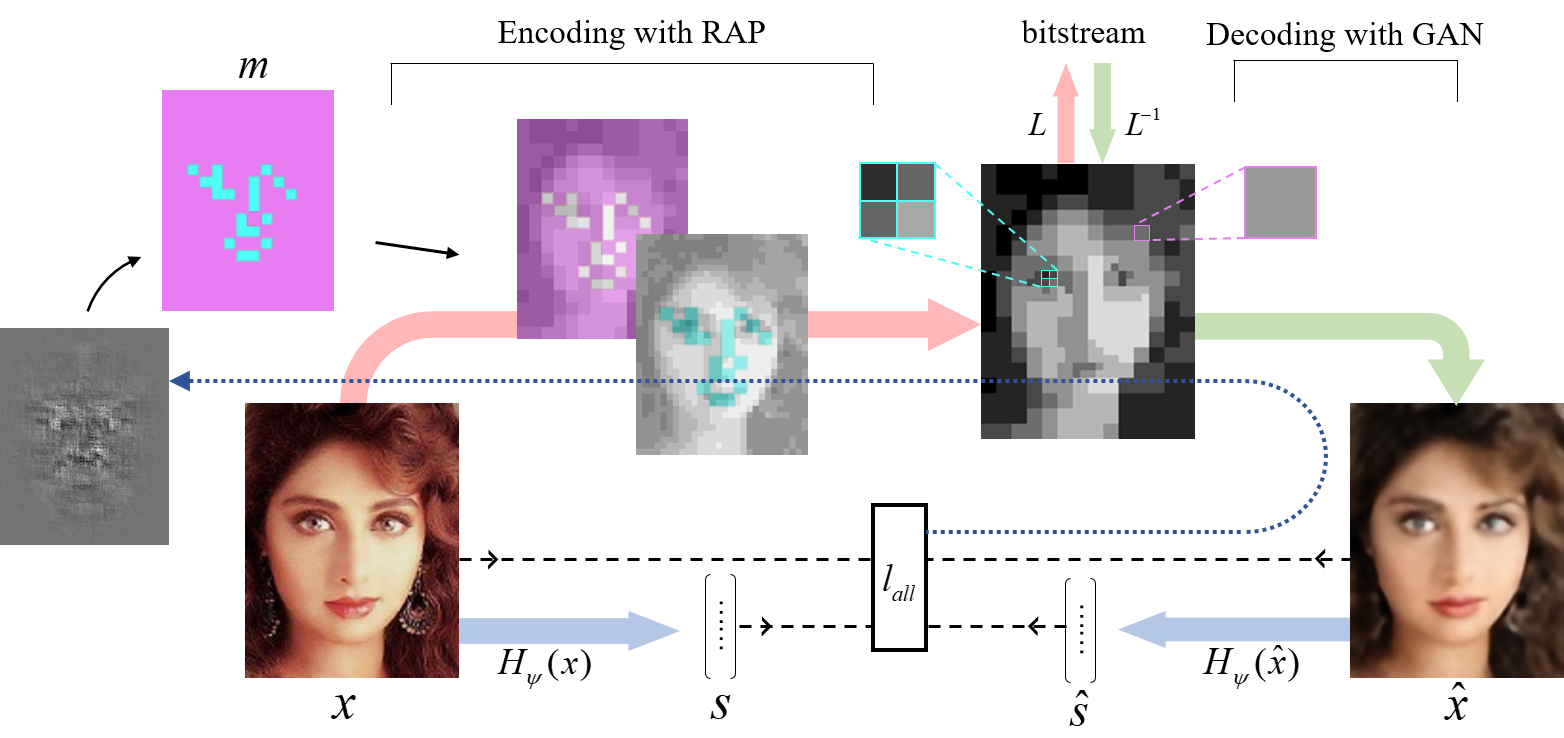}}
	\caption{Illustration of the facial image compression scheme adopted in this paper. The notations are consistent with Fig. \ref{fig:fig3}.}
	\centering
	\label{fig:fig4}
\end{figure*}

As mentioned in the introduction section, the proposed semantic-oriented facial image compression scheme incorporates the proposed pooling strategy, RAP (Regionally Adaptive Pooling), into the network, which is a differentiable and lossy operation that can use variable block size pooling for each image.

After RAP, We then implement a simple prediction as illustrated below:
\begin{equation}\label{6}
e^{(i,j)}=\left\{
\begin{aligned}
x^{(i,j)} \qquad \qquad \;, & \quad \text{if}\;i=1,j=1,  \\
x^{(i,j)}-x^{(i,j-1)}, & \quad \text{if}\;i=1,j>1, \\
x^{(i,j)}-x^{(i-1,j)}, & \quad \text{otherwise},
\end{aligned}
\right.
\end{equation}
where $(i,j)$ denote coordinates of pixels. The output of prediction $e$ is followed by an arithmetic coding. As we showed previously, we merge a transformation $H_\psi$ to send back error measured in semantic domain, which will be illustrated with more details in next section.


\subsection{Encoding with RAP}

Pooling layer is commonly employed to down-sample an input representation immediately after convolutional layer in neural networks, on the assumption that features are contained in the sub-regions. In most of the widely used neural network structures, the pooling blocks are not overlapped and fixed block size is used in each image. However, in the context of image compression, such fixed block size pooling scheme is improper in the case of heterogeneous texture distribution. In order to address this issue and increase the flexibility, we propose RAP of using variable block sizes pooling scheme. The choice of block sizes used in each sub-region are represented as a mask.

Suppose an input image $x\in\mathbb{R}^{I\times J\times K}$, where $I, J, K$ denote the height, width, channel of $x$ respectively. The output of non-overlapping pooling operation with fixed block size $n$ can be denoted as $x_{(n)}$, where $x_{(n)}\in\mathbb{R}^{\frac{I}{n}\times \frac{J}{n}\times K}$. Then we interpolate $x_{(n)}$ to $x_n$, where $x_n\in\mathbb{R}^{I\times J\times K}$, and concatenate $x_n$ along the last dimension:
\begin{equation}\label{7}
x_{concat} = [x_1, \dots, x_n, \dots, x_N], \quad 1\leq n\leq N,
\end{equation}
where $x_{concat} \in \mathbb{R}^{I \times J \times (K \times N)}$, and $N$ indicate the maximum block size. We define a mask $M\in[0, 1]^{I\times J\times (K \times N)}$, the output of RAP can be formulated as:
\begin{equation}\label{8}
x_{RAP}^{(i,j)}= \sum_{k=1}^{K \times N}(x_{concat}^{(i,j,k)} \times M^{(i,j,k)}), \, s.t. \sum_{k=1}^{K \times N}M^{(i,j,k)}=K,
\end{equation}
where $i,j,k$ denote indexes of height, width, channel respectively.

\begin{algorithm}[t]
	\caption{Updating Scheme of $M$}\label{alg}
	\begin{algorithmic}[1]
		\Procedure{Updating of $M$ at encoding time}{}
		\State $\textit{count} \gets 0$
		\State $\textit{budget} \gets \text{a given budget for the current image}$
		\State $\textit{max} \gets \text{maximum number of allowed loops}$
		\State $M \gets \text{initialization according to Equ. \ref{100}}$
		\BState \emph{top}:
		\If {$\textit{bitrate} > \textit{budget}$ or $\textit{count} > \textit{max}$} \Return false
		\EndIf
		\BState \emph{loop}:
		\State $G=\frac{dl_{all}}{dx_{RAP}}$
		\State $\{(i,j)\}=argmax[sum(G^{(i,j)})\text{ for each block}]$
		\For {(i,j)\text{ in }\{(i,j)\}}
		\State $M^{(i,j,index(M=1))}=0$
		\State $M^{(i,j,index(M=1)-1)}=1$
		\EndFor
		\State \textit{count} = \textit{count} + 1
		\State \textbf{goto} \emph{top}.
		\EndProcedure
	\end{algorithmic}
\end{algorithm}

In the training stage, the mask $M$ is random initialized to facilitate a robust learning process of neural networks. In the testing stage, at encoding time, $M$ is adaptively determined by: 1) a given bit rate budget; 2) the gradient feedback from the integrated semantic distortion metrics. We first initialize $M$ as:

\begin{equation}\label{100}
M^{(i,j,k)}=\left\{
\begin{aligned}
1, & \quad \text{if}\;k\geq K\times (N-1),  \\
0, & \quad \text{otherwise}, \\
\end{aligned}
\right.
\end{equation}
then automatically update $M$ according to Alg. \ref{alg}. 
In practice, we first set a constraint on the mask according to bit rate budget, then adjust the mask based on gradient feedback. For example, smaller block size will be used in the location determined by gradient feedback if the bit rate budget is adequate.
We encode $M$ with arithmetic coding as overhead (around 5\%-10\% of the total bit rate).

At decoding time, since the mask $M$ (transmitted as overhead) is available, we can completely restore the compact representation $c$ after arithmetic decoding. Finally, the reconstructed image is obtained by the decoder $G_\phi$.

Different from autoencoder-based compression, RAP provide the ability of preserving crucial local features that have great impact on face verification (\textit{e.g.}, the regions around eyes have larger gradient so that these regions should be pooled with smaller block size), which adds support of spatially adaptive bit allocation to our LFIC framework. We also believe that RAP has the potential to be embedded in the widely used convolutional neural network structure to provide the strong flexibility. The experimental results demonstrate that RAP served as a promising encoder component, and the restored faces retain the semantic property very well.

\subsection{Decoding with Adversarial Networks}

As a fast-growing architecture in the field of neural network, GANs achieve impressive success in lots of tasks. We apply such generative model to compression directly to reduce reconstruction error. We employ a discriminator $D_{\pi}$ training with decoder simultaneously, to force the decoded images to be indistinguishable from real images and to make the reconstruction process well constrained by incorporating prior knowledge of face distribution. Since standard procedures of GAN usually result in mode collapse and unstable training \cite{mao2016least}, therefore, we adopt the least square loss function in LSGAN \cite{mao2016least}, which yields minimizing the Pearson $\chi^2$ divergence instead of Jensen-Shannon divergence \cite{goodfellow2014generative}. Our adversarial loss can be defined as follows:
\begin{equation}\label{9}
l_{adv}=\frac{1}{2} \mathbb{E}_{x \sim P_{data}(x)} [(D_\pi (G_\phi(Q_\varphi(F_\theta(x)))) - 1)^2],
\end{equation}

Adversarial losses can urge the reconstructed data distributed as the original one in theory. However, a network with large enough capability can learn any mapping functions between these two distributions, that cannot guarantee the learned mapping producing desired reconstructed images. Therefore, a constraint to mapping function is needed to reduce the space of mapping functions. This issue calls for the employment of pixel-wise L1 loss for content consistency:
\begin{equation}\label{10}
l_{con}=\mathbb{E}_{x \sim P_{data}(x)} [{\lVert G_\phi(Q_\varphi(F_\theta(x))) - x \rVert}_1],
\end{equation}

\begin{table*}[h]
	\renewcommand\arraystretch{1.2}
	\centering
	\caption{Details of our decoder architecture. Each convolutional layers are optionally followed by several RBs (Residual Blocks).}
	\label{archi}
	\begin{tabular}{lcccccc}
		\toprule[2pt]
		\multicolumn{1}{c}{\multirow{2}{*}{Layer}} & \multirow{2}{*}{RBs} & \multirow{2}{*}{Input} & \multicolumn{1}{l}{Filter Size /} & \multirow{2}{*}{BN} & \multirow{2}{*}{Activation} & \multirow{2}{*}{Output} \\
		\multicolumn{1}{c}{}                       &                      &                        & Stride                            &                     &                             &                         \\ \hline
		1                                      & 1                    & input                  & $5\times5$ / 2                    & Y                   & ReLU                        & conv1    \\
		2                                      & 2                    & conv1                  & $3\times3$ / 2                    & Y                   & ReLU                        & conv2    \\
		3                                      & 2                    & conv2                  & $3\times3$ / 2                    & Y                   & ReLU                        & conv3   \\
		4                                      & 3                    & conv3                  & $3\times3$ / 2                    & Y                   & ReLU                        & conv4     \\
		5                                    & 2                    & conv4                  & $3\times3$ / 2                    & Y                   & ReLU                        & deconv5   \\
		6                                    & 2                    & deconv5, conv3         & $3\times3$ / 2                    & Y                   & ReLU                        & deconv6   \\
		7                                    & 1                    & deconv6, conv2         & $3\times3$ / 2                    & Y                   & ReLU                        & deconv7    \\
		8                                    & -                    & deconv7, conv1         & $3\times3$ / 2                    & Y                   & ReLU                        & deconv8  \\
		9                                      & 1                    & deconv8, input         & $3\times3$ / 1                    & Y                   & ReLU                        & conv9  \\
		10                                     & 1                    & conv9                  & $3\times3$ / 1                    & Y                   & ReLU                        & conv10  \\
		11                                     & -                    & conv10                 & $3\times3$ / 1                    & N                   & Tanh                        & conv11   \\
		\bottomrule[1.5pt]
	\end{tabular}
\end{table*}

\begin{table*}[h]
	\renewcommand\arraystretch{1.2}
	\centering
	\caption{Detailed architecture of RB (Residual Block)}
	\label{resblock}
	\begin{tabular}{llccccl}
		\toprule[2pt]
		\multicolumn{1}{c}{\multirow{2}{*}{Layer}} & \multicolumn{1}{c}{\multirow{2}{*}{Type}} & \multicolumn{1}{c}{\multirow{2}{*}{Input}} & Filter Size /  & \multirow{2}{*}{BN} & \multirow{2}{*}{Activation} & \multicolumn{1}{c}{\multirow{2}{*}{Output}} \\
		\multicolumn{1}{c}{}                       & \multicolumn{1}{c}{}                      & \multicolumn{1}{c}{}                       & Stride         &                     &                             & \multicolumn{1}{c}{}                        \\ \hline
		1                                          & Conv                                      & input                                      & 3$\times$3 / 1 & Y                   & ReLU                        & conv1                                       \\
		2                                          & Conv                                      & conv1                                      & 3$\times$3 / 1 & N                   & ReLU                        & conv2                                       \\
		3                                          & Add                                       & input, conv2                               & -              & -                   & -                           & output                                      \\
		\bottomrule[1.5pt]
	\end{tabular}
\end{table*}

\textbf{Decoder Architecture.} Previous works \cite{he2016deep} have shown that residual learning have the potential to train a very deep convolutional neural network. We employ several Convolution-BatchNorm-ReLU modules \cite{ioffe2015batch} and residual modules based on symmetric skip-connection architecture for the decoder, which allowing connection between a convolutional layer to its mirrored deconvolutional layer (Tab. \ref{archi}). Any extra inputs are specified in the \textit{input} column. Such design mix the information of different features extracted from various layers, and prevent training from suffering from gradient vanishing. For the discriminator network, we follow DCGAN \cite{radford2015unsupervised} except for the least square loss function.

\subsection{Training with Semantic Distortion Metric}
\label{facesemanticmetric}

Our main goal is to obtain a compact representation, and ideally, such representation is expressive enough to rebuild data under semantic distortion metric. As we have shown previously, each in-loop operation in our framework is differentiable to guarantee that the error will be propagated back. 

As to facial compression, we select FaceNet \cite{schroff2015facenet} as the metric transformation $H_\psi$, a neural-network-based tool that maps face images to a compact Euclidean space. Such space amplifies distances of faces from distinct people, while reduce distances of faces from the same person. This model is pre-trained with triplet loss and center loss \cite{wen2016discriminative}.
Specifically, we adopt L2 loss to facilitate semantic preserving on encoder and decoder:
\begin{equation}\label{11}
l_{sem}=\mathbb{E}_{x \sim P_{data}(x)} [{\lVert H_\psi(G_\phi(Q_\varphi(F_\theta(x)))) - H_\psi(x) \rVert}_2^2],
\end{equation}

\subsection{Full Objective}

The overall objective is a weighted sum of three individual objectives:
\begin{equation}\label{12}
l_{all}= \lambda_{con} l_{con} + \lambda_{sem} l_{sem} + \lambda_{adv} l_{adv},
\end{equation}

In practice, we also attempted a regularization term $l_{reg}$ and replace L1 norm with MSE, but did not observe obvious performance improvement.

\section{Experiments}
\label{experiment}

\begin{figure*}[t]
	\centerline{\includegraphics[width=\linewidth]{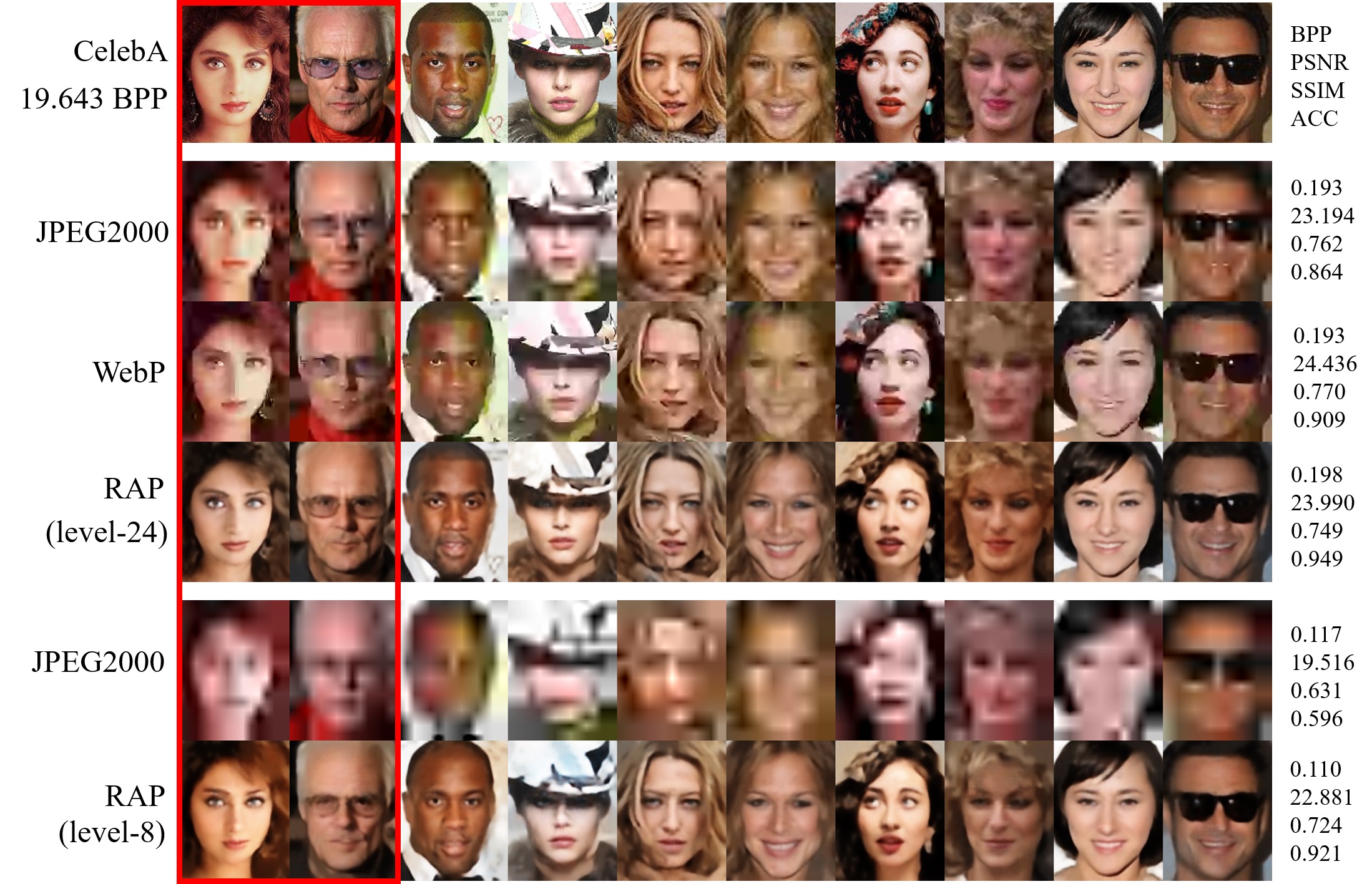}}
	\caption{Qualitative results of our model (RAP) compared to JPEG2000, WebP. Each value is averaged over testing set. ACC means the accuracy of face verification. Our results are obtained by training with 24 and 8 quantization levels respectively. The WebP codec can't compress images to a bit rate lower than 0.193 BPP. It is worth noting that as demonstrated in the first two columns, our scheme tend to learn the key facial structure instead of color of the girl's hair and the elder man's cloth.}
	\centering
	\label{fig:fig6}
\end{figure*}

In this section, we will introduce the dataset and the specific experimental details for facial compression. We compared our proposed method with traditional codecs and specific facial image compression methods. The results demonstrate that our method not only produce more visually preferred images under very low bit rate, but also good at preserving semantic information in the context of face verification scenario.

\textbf{Dataset.} We used the publicly available CelebA aligned \cite{liu2015deep} dataset to train our model. CelebA contains 10,177 number of identities and 202,599 number of facial images. We eliminated the faces that cannot be detected by dlib \footnote{http://dlib.net/}, and that are judged to be profiles based on landmarks annotations. The remaining images were cropped to $144\times112\times3$, and randomly divided into training set (100,000 images, 9014 identities) and testing set (14,871 images, 1870 identities).

\textbf{Evaluation.} We adopt the accuracy (10-fold cross-validation) of face verification to represent the ability of semantic preservation, which means that lower verification accuracy represents higher semantic distortion during image compression progress. Face verification is a binary classification task that given a pair of images, to determine whether the two pictures represent the same individual. We randomly generate 6000 pairs in testing set for face verification, where the positive and negative samples are half to half. The bitstream was represented as Bit-Per-Pixel (BPP). The Peak Signal-to-Noise Ratio (PSNR) was calculated in RGB channel. Furthermore, in order to calculate the equivalent rate distortion difference between two compression schemes, we refer \cite{bjontegaard2001calcuation}, which is widely used in international image/video compression standard. The only difference in implementation is that we replace BPS (Bits Per Second) by BPP as rate index, and replace PSNR by face verification accuracy as distortion index.

\textbf{Pre-train for metric transformation.} In our task, the metric transformation plays an important role in bridging pixel domain and semantic domain, where the distortion is measured to provide gradient feedback. We employed FaceNet \cite{schroff2015facenet}, a learned mapping that translating facial images to a compact Euclidean space, where the distance representing facial similarity. We use the parameters pre-trained on MS-Celeb-1M dataset \footnote{https://github.com/davidsandberg/facenet}, with a competitive classification accuracy of 99.3\% on LFW \cite{huang2007labeled} and 97.5\% on our generated pairs. In the training stage of compression network, the parameters of metric transformation will be fixed to ensure the reliability of semantic distortion measurement.

\textbf{Implementation details.} We implement all modules using TensorFlow, with training executed on NVIDIA Tesla K80 GPUs. We employ Adam optimization \cite{kingma2014adam} to train our network with learning rate 0.0001. All parameters are trained with 20000 iterations (64 images / iteration), which cost about 24 GPU-hours. We heuristically set $\lambda_{con}=0.01$, $\lambda_{sem}=10$, $\lambda_{adv}=0.1$  in the experiments. The principle is to increase the weight of adversarial and semantic parts as much as possible, while avoiding disturbance to subjective quality of reconstructions.
We adopt the block size of 4 or 8 as an instance to demonstrate the effectiveness of our scheme. We have also conducted our experiments with different block sizes (e.g., 16 or 32), but it doesn't bring much rate distortion performance gain compared with current settings.

\begin{figure*}[t]
	\centerline{\includegraphics[width=0.9\linewidth]{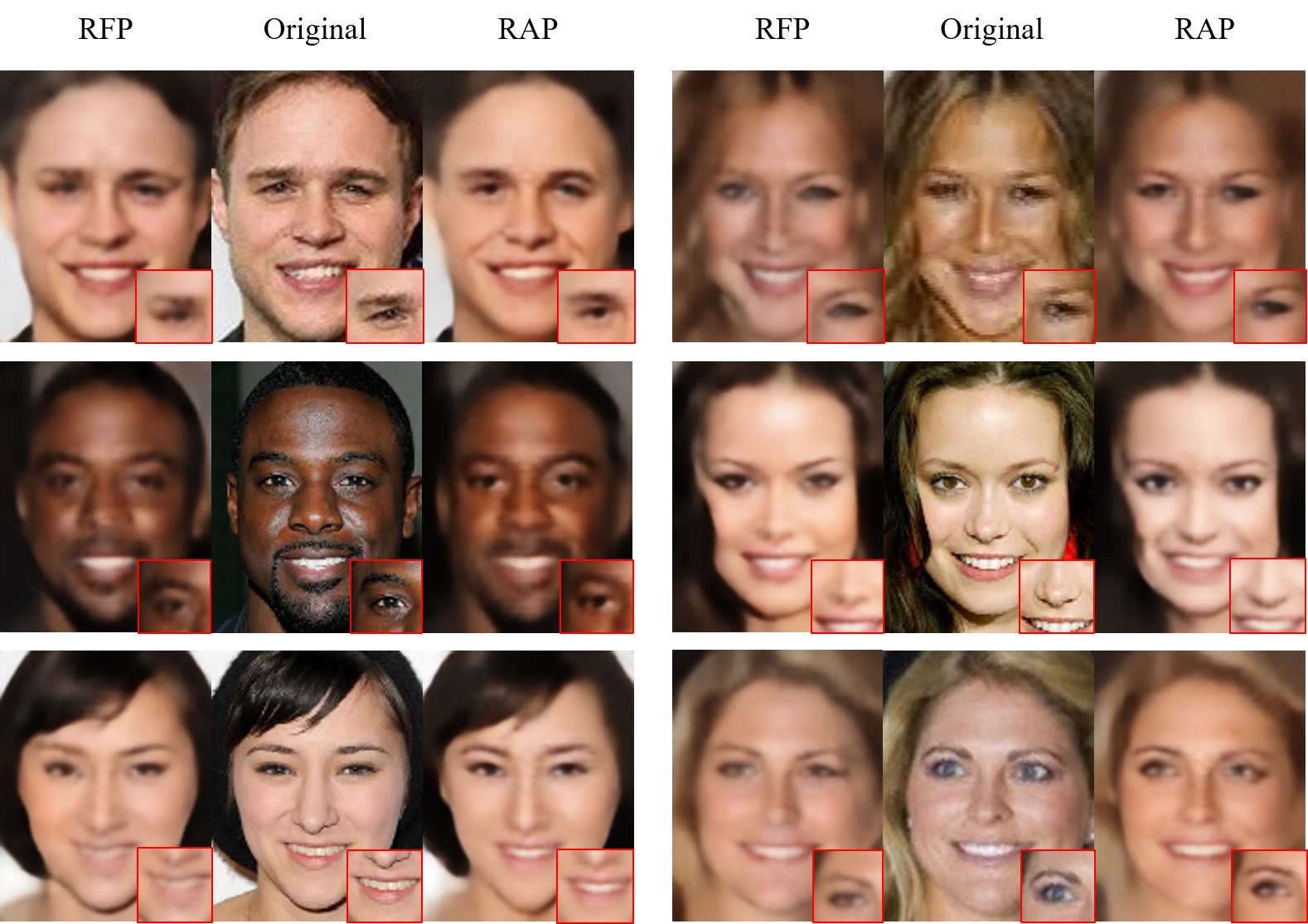}}
	\caption{Detailed comparison between RAP and RFP. The first and the fourth columns are decoded images from RFP at about $0.079$ BPP, while the third and the sixth columns are decoded images from RAP at about $0.073$ BPP. Obviously, RAP demonstrates much better performance in preserving distinguishable details than RFP.}
	\centering
	\label{fig:fig8}
\end{figure*}

\subsection{Comparison against Typical Image Compression Schemes}

We compare our model against typical widely used image compression codecs (JPEG2000 \footnote{http://www.openjpeg.org/ (v2.1.2)}, WebP \footnote{https://developers.google.com/speed/webp/ (v0.6.0-rc3)}), a neural network-based image compression method of Toderici \textit{et al.} \footnote{https://github.com/tensorflow/models/tree/master/research/compression} \cite{toderici2016full}  and specific facial image compression methods \cite{elad2007low,bryt2008compression,ram2014facial,ferdowsi2015sparse}. 

We refer \cite{bjontegaard2001calcuation} to calculate the equivalent rate distortion difference between two compression schemes. We replaced BPS by BPP, which is averaged over testing set. We also replaced PSNR by face verification accuracy, which is described in aforementioned section. The results outperform JPEG2000 and WebP codecs, as well as Toderici's solution significantly, as shown in Table.\ref{rate-distortion}.

Some considered efforts have been made to specific facial image compression \cite{elad2007low,bryt2008compression,ram2014facial,ferdowsi2015sparse}. But instead of automatically optimizing with integrated hybrid metrics (\textit{e.g.}, semantic fidelity), they adjusted compression parameters heuristically (\textit{e.g.}, bit allocation) and evaluated their performance of methods on gray-scale images with PSNR/SSIM only (34.20\% $\sim$ 47.18\% bit rate reduction over JPEG2000 as Tab. \ref{baseline} demonstrates \footnote{A fixed header size of 100 bytes in JPEG2000 is added for all results.}, these results are extracted from their papers since the authors don't release their source code for comparison).

\def\arraystretch{1}
\begin{table}[h]
	\centering
	\caption{Ratio of Bit Rate Saving of our scheme compared with benchmarks}
	\label{rate-distortion}
	\begin{tabular}{lc}
		\hline
		\diagbox{Anchor}{Test}         & Ours      \\ \hline
		JPEG2000 & -71.41\% \\
		WebP     & -48.28\% \\
		Toderici \textit{et al.} \cite{toderici2016full} & -52.67\%
	\end{tabular}
\end{table}

\def\arraystretch{1}
\begin{table}[h]
	\centering
	\caption{Comparison with specific facial image compression methods on ratio of Bit Rate Saving relative to JPEG2000}
	\label{baseline}
	\begin{tabular}{lc}
		\hline
		\diagbox{Test}{Anchor}         & JPEG2000      \\ \hline
		Elad \textit{et al.} \cite{elad2007low} & -47.18\% \\
		Bryt \textit{et al.} \cite{bryt2008compression}     & -45.55\% \\
		Ram \textit{et al.} \cite{ram2014facial} & -34.20\% \\
		Ferdowsi \textit{et al.} \cite{ferdowsi2015sparse} & -35.22\% \\
		Ours & -71.41\%
	\end{tabular}
\end{table}

As mentioned in introduction section, pixel fidelity cannot fully reflect semantic difference. For instance, in Fig. \ref{fig:fig6}, we can observe that RAP at 0.110 BPP has a much higher face verification rate and better visual experience than JPEG2000 and WebP at 0.193 BPP, even though RAP has lower PSNR/SSIM in this case. On the other hand, Delac \textit{et al.} \cite{delac2008image} found out that most traditional compression algorithms have face verification rate dropped significantly under the bit rate range of 0.2 $\sim$ 0.6 BPP. In contrast, our scheme can keep face verification accuracy without significantly deterioration even under the very low bit rate of 0.05 BPP.

We calculate the time consuming of our scheme and traditional codecs on the same machine (CPU: i7-4790K, GPU: NVIDIA GTX 1080). The overall computational complexity of our implementation is about $13$ times that of WebP. It should be noted that our scheme is just a preliminary exploration of learning-based framework for image compression and each part is implemented without any optimization.

\begin{figure}[h]
	\centerline{\includegraphics[width=0.7\linewidth]{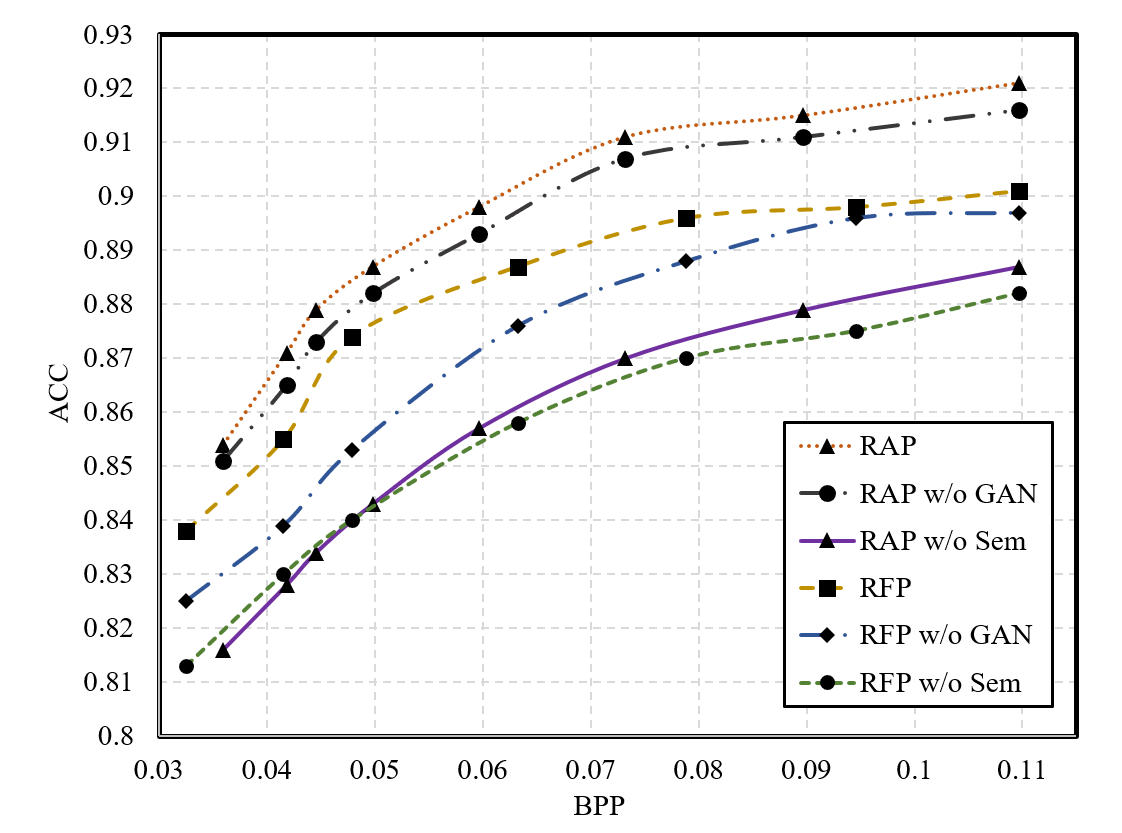}}
	\caption{Rate Distortion performance analysis. Cubic spline interpolation is used for fitting curves from discrete points.}
	\centering
	\label{fig:fig7}
\end{figure}

\subsection{Rate Distortion Performance Analysis}

To evaluate the effectiveness of spatially adaptively bit allocation, we compared RAP with its non-adaptive counterpart, namely, Regionally Fixed Pooling (RFP), whose block sizes are all fixed. Acting in this way, RFP could not adjust block sizes to achieve variable rate at testing time as RAP does. Therefore, we trained RFP model with different quantization step to realize variable rate for comparison.
As Fig. \ref{fig:fig7} illustrates, with the increase of bit budget, the performance of RAP is much higher than RFP. We also provide detailed comparison in Fig. \ref{fig:fig8} which demonstrates that RAP can automatically preserving better quality than RFP on the distinguishable regions under the same bit rate.

We also analyze the influence of adversarial loss $l_{adv}$ and semantic loss $l_{sem}$ by comparing the performance of RAP/RFP, RAP/RFP without adversarial loss (RFP w/o GAN) and RAP/RFP without semantic loss (RFP w/o Sem). We observed that both of these losses make contributions to our delightful results, and the semantic loss shows much higher influence than adversarial loss. Note that RAP without $l_{sem}$ is worse than RFP due to its failure to allocate more bits on distinguishable regions as illustrated in Fig. \ref{fig:fig2}.

\section{Conclusion}
\label{conclusion}

We introduce a LFIC framework integrated with the proposed Region Adaptive Pooling module and a general semantic distortion metric calculation module for task-driven facial image compression. The LFIC enables the image encoder to automatically optimize codec configuration according to integrated semantic distortion metric in an end-to-end optimization manner. Comprehensive experiment has been done to demonstrate the superior performance on our proposed framework compare with some typical image codecs and specific image codecs for facial image compression. We expect to refine prediction and entropy coding modules to further improve compression performance and apply the framework to more general scenarios in future work.

\section{Acknowledgement}
\label{acknowledgement}

This work was supported in part by the National Key Research and Development Program of China under Grant No. 2016YFC0801001, the National Program on Key Basic Research Projects (973 Program) under Grant 2015CB351803, NSFC under Grant 61571413, 61632001, 61390514.


\appendix


\section{More Experiments}

\begin{figure*}[h]
	\centerline{\includegraphics[width=0.8\linewidth]{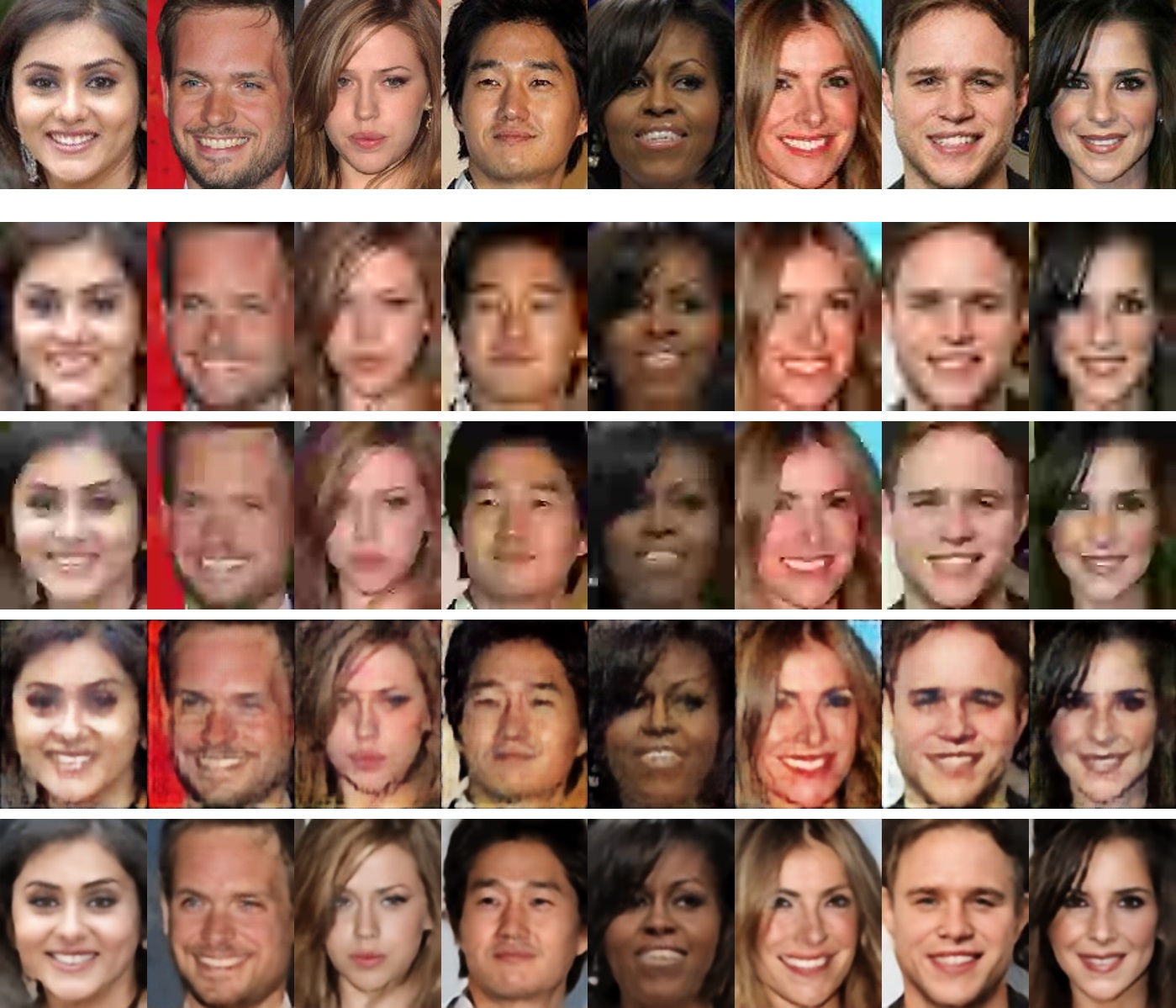}}
	\caption{The first row is uncompressed images. From the second row to the bottom, each row represents the decoded images from JPEG2000 ($0.193$ BPP), WebP ($0.193$ BPP), Toderici et al. ($0.250$ BPP) and RAP ($0.198$ BPP) respectively.}
	\centering
	\label{apx:fig1}
\end{figure*}

\begin{figure*}[h]
	\centerline{\includegraphics[width=0.8\linewidth]{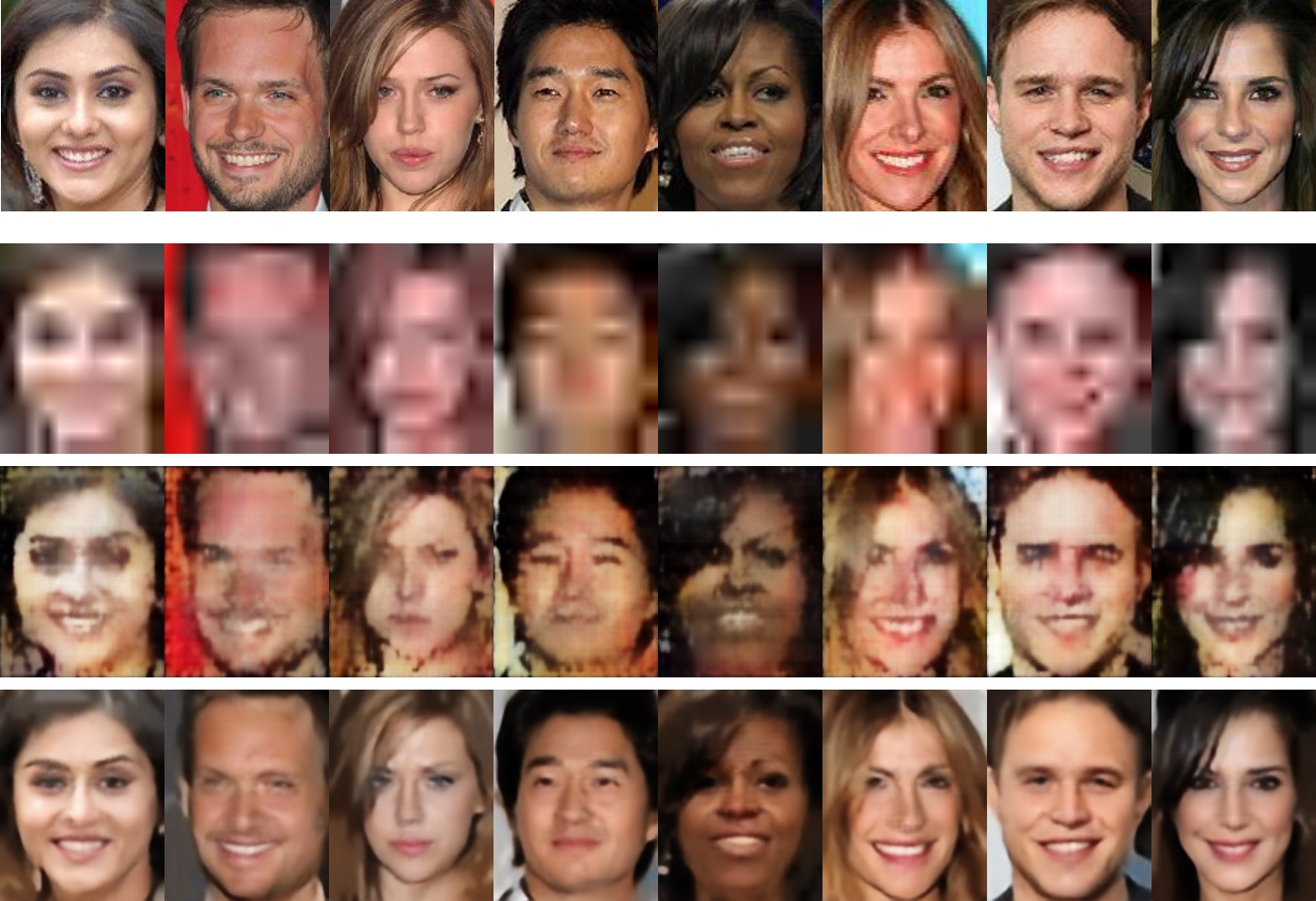}}
	\caption{The first row is uncompressed images. From the second row to the bottom, each row represents the decoded images from JPEG2000 ($0.117$ BPP), Toderici et al. ($0.125$ BPP) and RAP ($0.110$ BPP) respectively. The WebP codec can't compress images to a bit rate lower than $0.193$ BPP.}
	\centering
	\label{apx:fig2}
\end{figure*}

We give more qualitative results in Figure \ref{apx:fig1} and Figure \ref{apx:fig2}. Note that, as explained in paper, several specific facial image compression algorithms evaluated their performance using PSNR/SSIM only, and these algorithms' performance are ranged in $34.20$\% $\sim$ $47.18$\% bit rate reduction over JPEG2000, which are extracted from their published papers since the authors don't want to open their code for comparison. On the other hand, for most of specific facial compression algorithms, verification rate drops significantly under $0.2\sim0.6$ BPP. In contrast, our scheme does not deteriorate significantly even under $0.05$ BPP. We also compared our results with Toderici et al. \footnote{https://github.com/tensorflow/models/tree/master/research/compression} without entropy coding, since they didn't release their trained model for entropy coding.

To achieve this results, we integrate different metrics: adversarial loss, L1 loss and semantic loss. We leverage semantic metric to retain identity, while the use of content loss here is to constrain mapping between pixel space and semantic space. Accordingly, the weight of each metrics is a trade-off in this scheme and we heuristically adjust this hyperparameters at present. The principle is to increase the weight of semantic part as much as possible, while avoiding disturbance to subjective quality of reconstructions.

\cleardoublepage
\clearpage

\bibliographystyle{elsarticle-num-names}
\bibliography{face}

\end{document}